\documentclass[onecolumn,apj]{emulateapj}
\bibpunct{(}{)}{;}{a}{}{,} 
\usepackage{endnotes}
\usepackage{graphicx}
\usepackage{natbib}
\usepackage{txfonts}
\begin{document}
\title{Exploring Io's atmospheric composition with APEX: first measurement of $^{34}$SO$_2$ and tentative detection of KCl}
\author{A. Moullet\altaffilmark{1}}
\author{E. Lellouch\altaffilmark{2}}
\author{R. Moreno\altaffilmark{2}}
\author{M. Gurwell\altaffilmark{3}}
\author{J.~H Black\altaffilmark{4}}
\author{B. Butler\altaffilmark{5}}

\affil{\altaffilmark{1}National Radio Astronomy Observatory, Charlottesville VA-22902, U.S.A}
\affil{\altaffilmark{2}LESIA-Observatoire de Paris, 5 place J. Janssen, 92195 Meudon CEDEX, France}
\affil{\altaffilmark{3}Harvard-Smithsonian Center for Astrophysics, Cambridge MA-02138, U.S.A}
\affil{\altaffilmark{4}Department of Earth and Space Sciences, Chalmers University of Technology, Onsala Space Observatory, 43992 Onsala, Sweden}
\affil{\altaffilmark{5}National Radio Astronomy Observatory, Socorro NM-87801, U.S.A}

\begin{abstract}
The composition of Io's tenuous atmosphere is poorly constrained. Only the major species SO$_2$ and a handful of minor species have been positively identified, but a variety of other molecular species should be present, based on thermochemical equilibrium models of volcanic gas chemistry and the composition of Io's environment. This paper focuses on the spectral search for expected yet undetected molecular species (KCl, SiO, S$_2$O) and isotopes ($^{34}$SO$_2$). We analyze a disk-averaged spectrum of a potentially line-rich spectral window around 345~GHz, obtained in 2010 at the APEX-12m antenna (Atacama Pathfinder EXperiment). Using different models assuming either extended atmospheric distributions or a purely volcanically-sustained atmosphere, we tentatively measure the KCl relative abundance with respect to SO$_2$ and derive a range of 4$\times$10$^{-4}$-8$\times$10$^{-3}$. We do not detect SiO or S$_2$O and present new upper limits on their abundances. We also present the first measurement of the $^{34}$S/$^{32}$S isotopic ratio in gas phase on Io, which appears to be twice as high as the Earth and ISM reference values. Strong lines of SO$_2$ and SO are also analyzed to check for longitudinal variations of column density and relative abundance. Our models show that, based on their predicted relative abundance with respect to SO$_2$ in volcanic plumes, both the tentative KCl detection and SiO upper limit are compatible with a purely volcanic origin for these species.
\end{abstract}
%\authorrunning{A. Moullet et al.}
%\titlerunning{}
\maketitle
%\keywords{Planets and satellites: atmospheres}

%\maketitle

\section{Introduction}
The galilean moon Io holds a very tenuous atmosphere ($\sim$1-10~nbar pressure at the ground) that is unusual in many respects. This is the only known SO$_2$-dominated atmosphere ($\sim$90\% of the total pressure), and is ultimately dependent upon the volcanic activity of the moon, the most intense in the solar system.
The atmosphere acts as a reservoir feeding a plasma torus in orbit around Jupiter that sweeps the upper gas layers at a rate as high as 1~ton/second \citep{schneider2007}, thus requiring a continuous and efficient atmospheric replenishment mechanism. It is yet unclear whether the bulk of the day-side atmosphere is replenished directly by the outgassing of volcanic plumes or by sublimation of SO$_2$ volcanic condensates \citep[see review in][]{lellouch2005}. The latest results obtained on SO$_2$ spatial distribution \citep{feaga2009,moullet2010} and column density variation with heliocentric distance \citep{tsang2012} tend to support sublimation as the major immediate source.\\
Io's atmosphere displays marked longitudinal column density variations on the day-side, by a factor up to 10 along the equator \citep{spencer2005}. In addition, it is expected that the pressure collapses almost entirely on the night-side and at high latitudes, where the colder surface temperature cannot maintain a significant SO$_2$ column density through sublimation, except possibly around volcanic centers.\\
\\
Primarily due to instrumental sensitivity limitations the composition of Io's atmosphere has been poorly constrained observationally. Beyond SO$_2$ only three molecular species have been identified in the lower atmosphere: \\
- SO is the main expected photochemistry product of SO$_2$ and was first detected by \citet{lellouch1996} with an abundance of 3-10\% with respect to SO$_2$ (i.e., mixing ratio). Measurements of significant temporal variations in SO mixing ratio, as well as mapping of SO lines emission \citep{depater2007,moullet2010} that shows a spatial distribution more localized and linked to volcanic centers than the bulk SO$_2$ atmosphere, support the hypothesis that atmospheric SO is at least partially sustained by plume outgassing.\\
- NaCl has been detected by sub-millimeter spectroscopy \citep{lellouch2003} with a mixing ratio of 0.3-1.3\%, and could be the main carrier for the Na and Cl ions present in the plasma torus. This molecule has a very short lifetime in the atmosphere due to photolysis and rapid condensation on the surface \citep{moses2002}. The presence of NaCl in volcanic plumes had been predicted by thermochemical equilibrium models of volcanic gas chemistry \citep{fegley2000}. It is expected to be injected in the atmosphere directly through volcanic outgassing or sputtering of Na-bearing volcanic frosts by high-energy particles from Io's torus \citep{johnson1990}, but not sustained via sublimation. \\
%While a precise measurement of its abundance is difficult to obtain due to the lack of knowledge on its spatial distribution, the results of \citet{lellouch2003} allowed to constrain the Na/S and Cl/S atomic ratios in erupting magmas.\\
- S$_2$ was detected by HST over the Pele plume \citep{spencer2000} with a temporally variable \citep{jessup2004} yet significant mixing ratio (8-33\%), hinting at a primarily volcanic origin. As S$_2$ is expected to quickly condense on the surface, it should remain localized mostly around volcanic centers. On the ground, it could polymerize into S$_3$ and S$_4$, thought to be responsible for the observed ring-shaped red-colored deposits.\\
\\
In parallel to these observations, numerous theoretical models predicting the composition of Io's atmosphere have been developed. \citet{summers1996} showed that SO$_2$ photochemistry should produce highly volatile, poorly condensible molecular species in the lower atmosphere such as O$_2$, SO, S$_2$, as well as atomic species S and O, that, through atmospheric transport, can build-up in the night-side where O$_2$ and SO may be the main constituents. Surface sputtering could also trigger the production of non- or partially-condensible atomic and molecular gases \citep[S, K, Na, NaS, Na$_2$O,...][]{wiens1997,chrisey1988,haff1981}.\\
An extensive literature focuses on detailed modeling of the composition of the gas mixture expelled by volcanic plumes, for a variety of temperature and pressure conditions in volcanic conduits, as well as different bulk magmatic atomic ratios, resulting in a wide range of expected species and abundances \citep[i.e., ][]{zolotov1998,fegley2000,moses2002,schaefer2004}. Assuming bulk magma composition based on Io's plasma torus and chondritic abundances, sulfur- (S$_2$, S$_3$, S$_2$O), carbon- (CO, CO$_2$, OCS), silicate- (SiO, Si), potassium- (KCl, K) and sodium-bearing molecules and atoms (NaCl, Na), among others, may be present in volcanic plumes at mixing ratios higher than 1$\times$10$^{-4}$. On the other hand, hydrogen-bearing molecules that are often found in terrestrial plumes (such as H$_2$O and H$_2$S) may not be present at significant levels as Io appears to be hydrogen depleted \citep{fegley2000b}. Many of the volcanic species expected on Io have a short lifetime in the atmosphere due to destruction by photochemical reactions and/or high condensibility on the cold ground ($\sim$90-130~K), thus forming condensate deposits around the plumes. These deposits can be partially re-injected back in the atmosphere by sputtering or, for the more volatile species (such as SO$_2$), through sublimation.   \\
\\
By relating measurements of the mixing ratios of different species to the above models, one can in principle better characterize volcanic regimes, the interactions of the atmosphere with its environment (surface and plasma torus), as well as check the validity of photochemical models. A more detailed knowledge of the chemical composition of Io's atmosphere, including the detection of expected yet undetected species, is essential for the understanding of the moon as a whole. The study of Io's spectrum in the sub(millimeter) wavelength range, which contains many rotational transitions of species that are or could be present in the atmosphere, is the most promising technique to pursue for the exploration of Io's chemistry, and is now more feasible with the recent developments of very sensitive instruments.\\ 
In this paper, we report on a spectroscopic exploration project carried out with the 12-m APEX antenna (Atacama Pathinder EXperiment) coupled with the APEX-2 heterodyne receiver. Spectra were obtained in the summer and fall of 2010, and targeted rotational transitions of confirmed molecules (SO$_2$, SO), as well as expected yet undetected species (KCl, SiO, S$_2$O) and isotopes ($^{34}$SO$_2$). We describe the data  and the models used to derive column densities and mixing ratios. We present the first measurement of the $^{34}$S/$^{32}$S isotopic ratio in gaseous SO$_2$, a tentative detection of KCl, upper limits on SiO and S$_2$O, and inferences on the role of volcanism for atmospheric replenishment.

\section{Observations}
The data analyzed in this paper were obtained on the APEX antenna (Atacama Pathfinder EXperiment), a 12~m diameter dish located on Llano de Chajnantor, at an altitude of 5100~m, in Chile \citep{gusten2006}. The observations were performed in August and October 2010, aggregating a total of $\sim$3.3~hours of on-source integration time (corresponding to $\sim$16.5 total hours of observation, taking into account time on the off position and overheads), with 1.4~hours on the leading hemisphere of Io (Eastern elongation, longitudes 0-180$^{\circ}$W), and 1.9~hours on the trailing hemisphere (Western elongation, longitudes 180-360$^{\circ}$W). The relevant observational parameters for each observation period are gathered in Table \ref{ephem}.\\
The single-sideband Swedish Heterodyne Facility Instrument \citep[SHeFI,][]{lapkin2008} was tuned to cover the 344.100-345.100~GHz and 346.430-347.430~GHz spectral windows. This setup was designed so as to target at once multiple rotational transitions of different species: two SO$_2$ transitions (346.523 and 346.652~GHz), two SO transitions (344.310 and 346.528~GHz), one KCl transition (344.820~GHz), one SiO transition (347.330~GHz), five $^{34}$SO$_2$  transitions (344.245, 344.581, 344.808, 344.987 and 344.998~GHz) and five S$_2$O transitions (344.851, 346.543, 346.804, 346.862 and 347.123~GHz). The Fast Fourier Transform Spectrometer backend \cite[FFTS,][]{klein2006} provided a spectral resolution of 122~kHz (equivalent to 106~m/s at 345~GHz), which is sufficient to resolve the Doppler-broadened lines ($\sim$600-800~m/s width). At the observed frequency, the antenna beam is $\sim$17" and does not resolve Io's disk ($\sim$1.2"), so that all the presented measurements are disk-averaged spectra.\\
\\
The observations were performed in on-off mode, hence in the raw spectra most of the sky thermal contribution is already removed, and the instrumental bandpass response corrected for. We noticed that the continuum level could vary significantly from scan to scan, and even reach large negative values, however line emission above the continuum is not affected by the continuum variations. These variations are probably related to the contribution of a strong continuum source (Jupiter) in the sidelobes. In the scans that are the most affected, the spectra can also display a large variability of the bandpass response on scales of a few hundred MHz (ripples). These issues prevent us from correctly estimating Io's continuum emission level, but still allow us to measure  the emission contrast of each line. To extract the spectra, using the GILDAS-CLASS package\footnote{http://www.iram.fr/IRAMFR/GILDAS/}, we carefully removed the continuum emission in each scan after fitting the continuum baseline by a polynomial (of order lower or equal to 2). A few scans where the continuum level could not be well estimated were flagged and removed. To account for Io's angular size variation from August to October 2010, we rescaled the October 2010 data by normalizing each scan's antenna temperature to the equivalent antenna temperature for the geocentric distance of August 2010. We then combined together the obtained scans by averaging them on a common rest frequency scale, taking into account the line-of-sight projected velocity of the source at the time of each scan.\\
The continuum-subtracted spectrum obtained is measured in the T$_A$$^{\star}$ scale, corresponding to the antenna temperature corrected for atmospheric absorption. To convert this scale to a disk-averaged brightness temperature scale, we multiplied the T$_A$$^{\star}$ values by the dilution factor (taking into account Io's ephemeris size and the beam size at each frequency), and by the antenna efficiency equal to B$_{eff}$/F$_{eff}$, where the telescope forward efficiency F$_{eff}$ is estimated at 0.95 and the main-beam efficiency B$_{eff}$ at 0.73 \citep{gusten2006}. We estimate an error of 5\% on the determination of the brightness temperature scale.\\
After combining all the obtained scans, the rms on the spectrum obtained (combined spectrum) is 9.5~mK at 344.8~GHz on the antenna temperature scale (for each spectral channel), corresponding to 3.8~K in equivalent disk-averaged brightness temperature on Io. When binned to obtain a spectral resolution of 488~kHz (424~m/s at 345~GHz), closer to the expected line-widths, the rms decreases to 4.8~mK (1.9~K in equivalent disk-averaged brightness temperature).

\begin{table}
\begin{center}
\begin{tabular}{|c|c|c|c|c|}
\hline 
Date & Angular size & Central longitudes & On-source time & Rms on spectrum \\
     & (")        & ($^{\circ}$W) & (h)  &  (mK/channel)\\
\hline
August 20, 2010 & 1.226 & 90-150 & 1.4 & 15\\
August 21, 2010 & 1.227 & 293-315 & 0.5 & 20 \\
October 13, 2010 & 1.247 & 279-312 & 0.7 & 23\\
October 15, 2010 & 1.242 & 300-329 & 0.7 & 23\\
Combined spectrum &  1.226 & & 3.3 & 9.5\\
\hline
\end{tabular}
\caption{Io's observational parameters at the time of observations. %From JPL Horizons ephemeris (http://ssd.jpl.nasa.gov/?horizons). The average projected radius (1821~km) is used to compute Io's angular size. % 
To obtain the combined spectrum, the antenna temperature scale of the individual scans is rescaled to the same apparent angular size, before averaging all the scans together}. The rms is determined by the uncertainty on the spectrum continuum baseline around 344.8~GHz, using the original 122~kHz spectral resolution.\label{ephem}
\end{center}
\end{table}

%\begin{table}
%\begin{center}
%\begin{tabular}{|c|c|c|c|}
%\hline
%Species & Frequency & Lower level energy & Intensity at 300~K \\
%        &  (GHz) & (cm$^{-1}$) &  (log)\\
%\hline
%SO$_2$ & 346.523 & 102.75 & -2.73\\
%    & 346.652 & 105.29 & -2.47\\
%$^{34}$SO$_2$ &  344.245 & 49.94  & -2.88\\
%& 344.581 &  104.86 & -2.48\\
%& 344.808 & 72.91 &  -2.79\\
%& 344.987 & 91.44 & -2.75 \\
%& 344.998 & 56.94 & -2.84 \\
%SO & 344.310 & 49.31 & -1.87 \\
%& 346.528 & 43.19 & -1.79\\
%KCl & 344.820 & 253.48 & -0.31\\
%SiO & 347.330 & 40.55 & -0.76\\
%S$_2$O & 344.851 & 301.29 & -3.87 \\
%& 346.543 & 216.65 & -3.43\\
%& 346.804 & 216.65 & -3.64\\
%& 346.862 & 216.63 & -3.64\\
%& 347.123 & 216.63 & -3.43\\
%\hline
%\end{tabular}
%\caption{List of targeted rotational lines. The spectroscopic parameters are retrieved from www.splatalogue.net and the CDMS database (http://www.astro.uni-koeln.de/cdms/catalog). Vibrational energy levels (not shown) are obtained from the Computational Chemistry Database (http://cccbdb.nist.gov/)\label{lines}}
%\end{center}
%\end{table}

\section{Atmospheric modeling}

\subsection{Radiative transfer}
To create synthetic atmospheric disk-averaged lines against which to compare the data, we use a numerical radiative transfer model described in detail in \citet{moullet2008}. For a given transition, considering the local gas temperature, density and velocity in each cell of a 3-D spatial grid encompassing the whole atmosphere, the code computes each local rotational line opacity profile, using spectroscopic parameters retrieved from the Splatalogue catalogue\footnote{www.splatalogue.net} and the CDMS database\footnote{http://www.astro.uni-koeln.de/cdms/catalog}. Pressure broadening of the line profile is neglected at the low pressures considered here, so that the line profiles are assumed to be only thermally and dynamically Doppler-broadened.  The brightness temperature corresponding to each line-of-sight direction is then computed through the corresponding atmospheric column, under the assumption that the emission is in local thermodynamical equilibrium. A continuum thermal surface emission at the brightness temperature of 95~K is assumed, corresponding to the average brightness temperature measured at $\sim$345~GHz \citep{moullet2010}. The synthetic disk-averaged line is finally derived by summing all local brightness temperatures in the grid and rescaling the result to Io's disk size.

\subsection{Extended atmospheric distributions (hydrostatic case)\label{extended}}
We first consider the case of an atmospheric global structure mimicking a horizontally extended atmosphere in hydrostatic equilibrium - akin to the case of a sublimation-sustained bulk atmosphere. The primary purpose of this modeling work is to determine the disk-averaged SO$_2$ column density and the abundances of the observed minor species with respect to SO$_2$ (i.e., mixing ratios).\\
We use a computing grid with 30~km resolution in the plane of sky, and 0.25~km resolution in altitude. The airmass at each cell is computed assuming plane parallel geometry. Our atmospheric distribution model assumes that SO$_2$ and all other species are co-located vertically and horizontally. Emission on the limbs, corresponding to the terminator region, is ignored since sublimation is unlikely to support significant amounts of SO$_2$ on the night-side. Otherwise, we assume either a horizontally constant column density {\it d} (homogeneous distribution model) or a spatially variable column density, taking as a reference the atmospheric distribution model proposed by \citet{feaga2009} based on HST disk-resolved SO$_2$  observations. The latter model proposes an atmosphere restricted to longitudes lower than 70$^{\circ}$, with column densities increasing towards the equatorial region. It also displays significant longitudinal variations (from 1$\times$10$^{16}$mol.cm$^{-2}$ to 5$\times$10$^{16}$mol.cm$^{-2}$ at the equator), with lower column densities on the sub-jovian hemisphere than on the anti-jovian hemisphere. We utilize the Feaga model distribution as a reference for relative column density variations across the disk, but allow the overall distribution to be scaled by a single parametric factor $\chi$.\\ For both homogeneous and Feaga distribution models, the vertical density profile within each atmospheric column is computed assuming hydrostatic equilibrium, using the SO$_2$ scale height for all species. We assume that the atmospheric temperature {\it T} is uniform over the entire atmospheric column, and horizontally constant across the disk. Doppler-shifts produced by atmospheric winds (and Io's solid rotation of 75~m/s at the equator) are also included. While most dynamic models of Io's atmosphere predict a day-to-night global circulation driven by pressure gradients \citep{ingersoll1989,austin2000}, the only detection of planetary winds on Io showed a structure similar to a prograde zonal wind, that can also conveniently explain the observed width of rotational lines \citep{moullet2008}. We hence introduce in our model a prograde zonal wind, characterized by the value of its equatorial velocity {\it V}, and for which, similarly to a solid rotation, velocity varies as a function of latitude as $\cos(lat)$. The projection of the zonal wind velocity on the line of sight is computed for each atmospheric column.\\
We use the minimum $\chi^{2}$ method on a grid of parameters to estimate the best set of atmospheric parameters ({\it d} or $\chi$, {\it T}, {\it V}) that can reproduce each observed line. The line contrast is mainly dependent on {\it T} and {\it d}, and, if the lines are not saturated, the Doppler line-width is mostly controlled by {\it T} and {\it V}. To break the possible degeneracy between retrieved parameters, when possible, we fit together different lines corresponding to distinct rotational transitions of the same molecule. Indeed, for a given molecule, the relative contrast between two lines of different intrinsic strength is mostly related to column density {\it d}, thus allowing for a quasi-independent determination of {\it d}. Here, by fitting alltogether the absolute and relative line contrasts and widths on the two strong SO$_2$ lines at 346.523 and 346.652~GHz, we obtain an optimal solution for {\it $d_{SO_2}$} (or scaling parameter $\chi$ in the case of non-uniform distribution), allowing for a quasi-independent determination of {\it T} and {\it V}. Assuming that all species share the same {\it T} and {\it V} parameters and are co-located with SO$_2$, we can then independently adjust for each minor species disk-averaged column density (and hence mixing ratio {\it q}).  \\

\subsection{Volcanically-sustained case\label{volc}}
To investigate the possible role of volcanism for minor species (SO, KCl, SiO and S$_2$O), we use a different atmospheric model than the one described above, based on realistic models of gaseous plumes in a rarefied environment developed by \citet{zhang2003} to simulate a plume-sustained atmosphere \citep[see details in][]{moullet2008}. These models do not include a background sublimation-sustained SO$_2$ component, so they only represent the direct volcanic contribution to the atmosphere. \\
The plume models fix the temperature and wind fields within a volcanic plume, as well as the spatial distribution of SO$_2$, corresponding to a canopy-shaped ballistic structure. The SO$_2$ total content in the plume, that ultimately relates to the SO$_2$ production rate from the vent, is also fixed so as to match SO$_2$ column densities measured over plumes by \citet{mcgrath2000} and \citet{jessup2004}. Two types of plumes are considered corresponding to those identified on Io \citep{mcewen1983}: Pele-type plumes, which are very large ($\sim$600~km radius) and rather rare - and the more common Prometheus-type ($\sim$200~km radius). Hence, the SO$_2$ emission from a given plume type (Prometheus or Pele) is entirely specified from the plume model (with no free parameters).
Within each plume, the only adjustable parameter is the minor species mixing ratio  {\it q} in the plume. By using a single {\it q} parameter for the whole plume, we hence assume that SO$_2$ and all other species share the same spatial distribution in the plume.\\
The total volcanic emission from an observed hemisphere can be modeled using a variable number of simultaneously active plumes of each category, placed at different locations on the disk. Our radiative transfer model allows simulation of any specified distribution of such plumes. In particular, we consider as the reference plume distribution the case where all plumes observed and localized by Galileo are active \citep{geissler2007}. This reference distribution includes 16 plumes, concentrated in particular on the anti-jovian side, two of which being Pele-type plumes while the others are Prometheus-type plumes. The corresponding volcanically-sustained atmospheric distribution hence exhibits large spatial variations of temperature and column density across the disk.\\
\\
Volcanic models are used here to fit the SO, KCl and SiO observations, using different plausible plumes distributions, ranging from the most favorable case (reference plume distribution), to a minimal case where only one small plume near the sub-earth point is active. For each plume distribution case, we determine the best fitting {\it q} parameter in plumes that would be needed to reproduce the observed line contrast (or the upper limit on line contrast). The range of retrieved {\it q} parameters is then compared, for each species, to the range of mixing ratios predicted by thermochemical equilibrium models of volcanic gas chemistry. The goal eventually is to determine what fraction of the minor species' atmospheric content can plausibly be directly sustained by active volcanic plumes.

\section{SO$_2$ and SO: longitudinal distribution}
\subsection{SO$_2$}
\begin{figure}
\begin{center}
\includegraphics{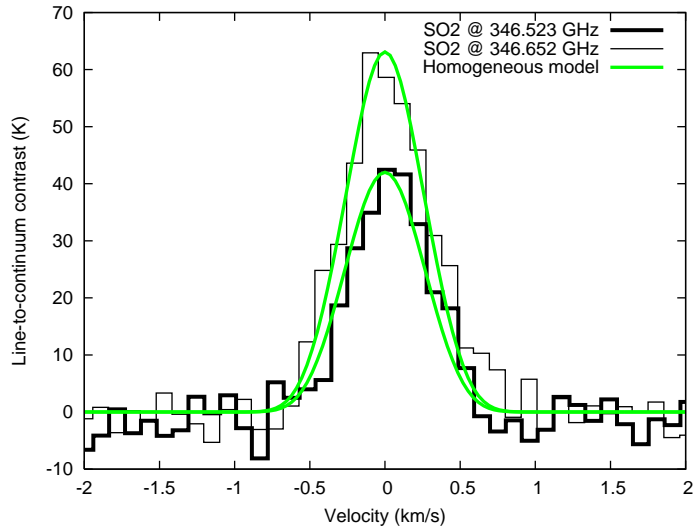}
\end{center}
\caption{SO$_2$ lines measured on the combined spectrum (average of all scans), plotted against the best synthetic line models for an homogeneous atmospheric distribution. \label{allSO2} }
\end{figure}

As measured on the combined spectrum (average of all obtained scans), the two SO$_2$ lines at 346.652 and 346.523~GHz offer a  peak signal-to-noise ratio (SNR) of 14 and 10 per spectral channel respectively. The best fitting parameters obtained with the extended atmosphere models are gathered in Table \ref{SO2models} (line 'combined'). For the homogeneous distribution model, we derive a column density {\it{d}}=5.5($\pm$0.7)$\times$10$^{15}$mol.cm$^{-2}$, a temperature {\it{T}}=220$\pm$20~K, and a wind velocity {\it{V}}=230$\pm$30~m/s. \\
The addition of a prograde wind appears to be necessary to reach a satisfactory fit, and the {\it{V}} values found are compatible with the direct wind measurement of \citet{moullet2008}. The {\it{T}} values retrieved are typical of findings from (sub)millimeter and UV spectral data \citep{jessup2004}, but higher than the upper limit derived from IR data \citep[$\sim$160~K,][]{tsang2012}. The data are also well reproduced for a column density distribution very similar to the Feaga model ($\chi$=0.9). We notice that the disk-averaged column density in our rescaled Feaga model is somewhat higher than the value for an homogeneous distribution (8.3 versus 5.5$\times$10$^{15}$mol.cm$^{-2}$), since the former contains higher-density regions in which lines are saturated. The best model for an homogeneous atmospheric distribution is plotted against the observed lines in Figure \ref{allSO2}. \\
\\
For these strong SO$_2$ lines, the quality of the data is sufficient to split the dataset into five (5) portions covering  $\sim$30$^{\circ}$ of rotational phase each (see Table \ref{SO2models}). The same modeling approach was performed on each portion in order to estimate atmospheric longitudinal variations. On these shorter datasets, uncertainties on the best fitted parameters reach $\sim$1.2$\times$10$^{15}$mol.cm $^{-2}$ on {\it{d}}, 40~K on {\it{T}}, and 50~m/s on {\it{V}}. Comparing the results obtained on different dates, it is not straightforward in this dataset to distinguish between longitudinal and temporal variations, as the data were obtained on two observation runs three months apart. The only direct temporal comparison can be done between the August 21 and October 13 datasets, that span approximately the same longitudes, and give compatible modeling results. \\
SO$_2$ disk-averaged column densities on both hemispheres appear to be of the same order, although they tend to be smaller on the trailing hemisphere than on the leading hemisphere. With scaling factors $\chi$ between 0.7-1.1, our rescaled Feaga models roughly follow the longitudinal column densities variations trend proposed by the original Feaga distribution model. We find the only significant departure from the original Feaga model between central longitudes 90$^{\circ}$W and 150$^{\circ}$W, where our best models propose a quasi-constant disk-averaged column density while the original Feaga model displays a $\sim$20\% increase.  \\
The prograde wind velocities appear to be systematically higher on the leading hemisphere than on the trailing hemisphere, where, at least for the case of rescaled Feaga models, the addition of a zonal wind is actually not required to provide a good fit. We note that a prograde zonal wind has been directly observed only on the leading hemisphere \citep{moullet2008}.\\
Temperatures tend to be lower on the trailing hemisphere than on the leading hemisphere, with the exception of the high temperatures obtained for the October 15th dataset. This particular dataset spans a large portion of the sub-jovian hemisphere where lower SO$_2$ column densities (by a factor 2-4) are measured, a predicted consequence of daily eclipses \citep{feaga2009,walker2012} which may make it more sensitive to plasma collisional heating \citep{wong2000}.

\begin{table}
\scriptsize

\begin{center}
\begin{tabular}{|c|c|c|c|c|c|c|c|}
\hline
\multicolumn{2}{|c|}{}&\multicolumn{3}{|c|}{Homogeneous model}&\multicolumn{3}{|c|}{Feaga model} \\
\hline
Date & Central longitudes &Column density & Temperature & Wind Speed &Scaling factor & Temperature & Wind Speed  \\
 & ($^{\circ}$W) & (10$^{16}$mol.cm$^{-2}$) & (K) & (km/s) & & (K) & (m/s)\\
\hline
Combined &  & 0.55 & 220 & 230 & 0.9 & 260 & 150\\
\hline
August 20, 2010 & 90-114 & 0.65 & 280 & 180 & 0.9 & 350 & 60\\
August 20, 2010 & 115-150 & 0.65 & 240 & 240 & 0.7 & 320 & 140\\
August 20, 2010 & 90-150  & 0.70 & 230 & 250  & 0.8 & 330 & 130 \\
August 21, 2010 & 293-315 & 0.50 & 190 & 110 &  1.1 & 220 & 40 \\
October 13, 2010 & 279-312 & 0.65 & 150 & 110 & 0.9 & 190 & 40\\
October 15, 2010 & 300-329 &  0.35 & 260 & 90 &  0.9 & 290 & 20\\
October 13-15, 2010 & 279-312/ 300-329 & 0.45 & 190 & 170 & 0.9 & 300 & 20\\
\hline
\end{tabular}
\caption{Best-fit SO$_2$ column density (or scaling factor $\chi$), temperature and zonal wind velocity assuming homogeneous and Feaga distribution models, for different portions of the dataset.  \label{SO2models}} 
\end{center}
\end{table}

\subsection{SO}
\begin{figure}
\begin{center}
\includegraphics{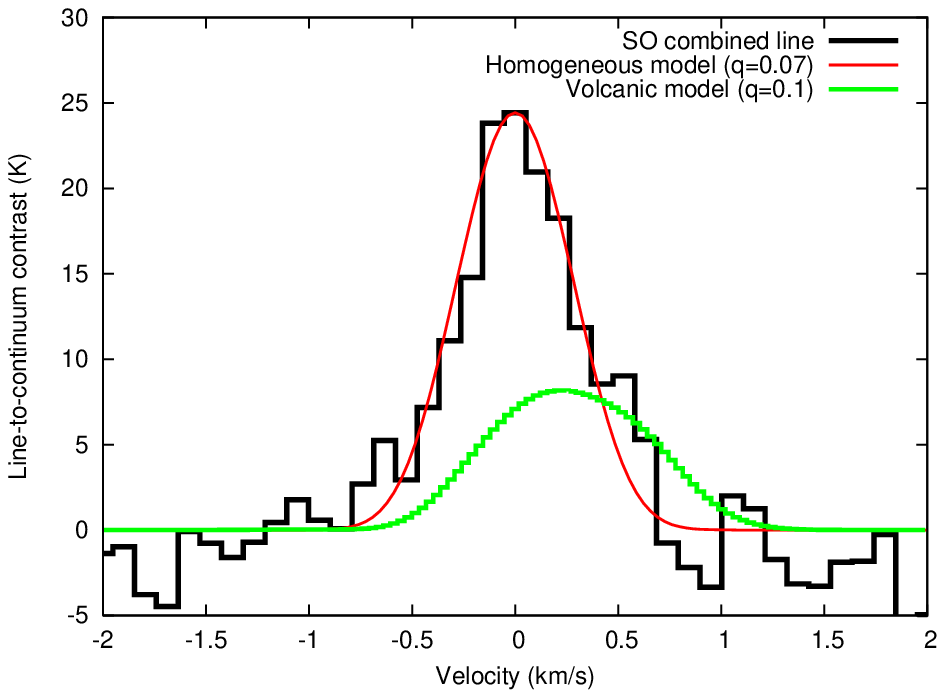}
\end{center}
\caption{Average of the two SO lines measured on the combined spectrum}, plotted against the best synthetic line model for an homogeneous atmospheric distribution and for a volcanic model assuming the reference plume distribution and q$_{SO}$=0.1. The volcanic model line appears to be slanted towards redshifted velocities, as it is dominated by the emission of plumes' infalling gas.\label{SOfig}
\end{figure}

To derive the SO mixing ratio, we average on a common velocity scale the 344.310 and 346.528~GHz SO lines measured on the combined spectrum, assigning an equal weight to each transition. On the resulting averaged line, a SNR of 9 per channel is reached on the line contrast (Figure \ref{SOfig}). We consider the atmospheric models discussed in Section \ref{extended} and assume that SO and SO$_2$ are co-located. For a rigorous comparison, the radiative transfer code is run for both SO transitions, and the two obtained synthetic SO lines are averaged before being compared to the average observed SO line.\\
We find the best line fit for a SO mixing ratio of q=0.07$\pm$0.007 (homogeneous distribution) or q=0.06$\pm$0.007 (rescaled Feaga distribution), corresponding to a disk-averaged SO column density of respectively 3.8 and 4.5$\times$10$^{14}$mol.cm$^{-2}$. Given the data quality, it is possible to split the data in 3 different portions - August 20, August 21 and October (13 and 15) - to look for mixing ratio variations. The values found (respectively q=0.06, 0.05 and 0.08 for the homogeneous case), given the error bars ($\sim$0.012), do not support significant mixing ratio variations, and are compatible with a scenario where SO is continuously present and detectable in the atmosphere.\\
These values fall in the mixing ratio range derived from previous (sub)mm observations \citep[q=0.03-0.1][]{lellouch1996,moullet2010}. The assumption on co-location between SO and SO$_2$ in our extended atmosphere models could correspond to the case of SO being entirely sustained by SO$_2$ photolysis, for which SO mixing ratios lower than 0.15 are indeed predicted \citep{summers1996}. However several arguments in favor of an additional volcanic source for SO have been proposed, including a more localized and plume-linked distribution for SO compared to SO$_2$ \citep{moullet2010} and the detection of temporal mixing ratio variations \citep{depater2007}.\\
While our disk-averaged spectrum does not carry information on SO spatial distribution, we can investigate the possibility of a purely volcanically sustained SO atmosphere with our volcanic models. As explained in section \ref{volc}, we assume plausible plume distributions, and derive the SO mixing ratio in plumes needed to reproduce the observed SO line. Assuming the most favorable case where all known plumes are active (reference plume distribution), the line contrast observed is reached for {\it q$_{SO}$}$=$0.7, which is a much larger value than the mixing ratios predicted from volcanic thermochemical models \citep[$<$10\%, e.g.,][]{zolotov1998}. Consequently, even higher SO mixing ratios would be required if fewer plumes were active. These results demonstrate that the SO content measured in our observations cannot be entirely sustained by plume activity, at least in the context of the latest thermochemical models. To determine an upper limit to the volcanic contribution to SO total content, we simulate the emission produced in the most favorable yet realistic volcanic case, by fixing the SO mixing ratio  in the plumes at q=10\% (corresponding to the upper range of predicted mixing ratios) and using the reference plume distribution. We then find that, within our assumptions on realistic SO abundance and plumes distribution, volcanic SO emission could only explain up to 30\% of the total emission observed. Hence our data are in agreement with the interpretation of previous (sub)mm observations, in which direct volcanic input of SO can only sustain a fraction of the SO content in the atmosphere, while SO$_2$ photolysis is the main source for SO.

\section{New detections and upper limits}
\subsection{First detection of gaseous $^{34}$SO$_2$}
\begin{figure}
\begin{center}
\includegraphics{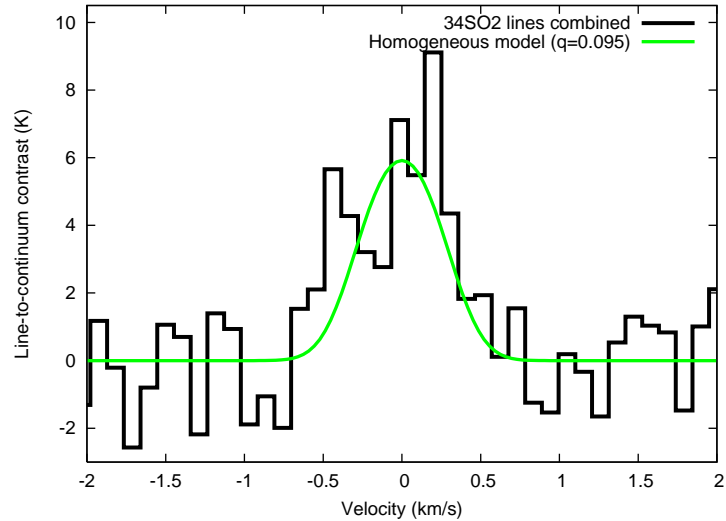}
\end{center}
\caption{Average of the five $^{34}$SO$_2$ lines measured on the combined spectrum, plotted against the best synthetic line models for an homogeneous atmospheric distribution.\label{34SO2fig}}
\end{figure}

While the most common isotope of sulfur, $^{32}$S, has been repeatedly detected both in solid and gaseous compounds, the sulfur isotopic composition has been only poorly constrained. A detection of the second-most common sulfur isotope ($^{34}$S) was only reported once, in a 4~$\mu$m $^{34}$SO$_2$ frost band \citep{howell1989}. However, the $^{34}$S/$^{32}$S ratio could not be derived, as the observed bands of $^{32}$SO$_2$ were too saturated to directly compare bands depths. \\
To attempt the first determination of the $^{34}$S/$^{32}$S ratio on Io, we averaged on a common velocity scale 5 different rotational lines of $^{34}$SO$_2$ that were each marginally detected on the combined spectrum, assigning an equal weight to each transition. The resulting average line is detected with a SNR of $\sim$4 on the line contrast (Figure \ref{34SO2fig}). In a similar way than for SO, the data is compared to the average of the same 5 $^{34}$SO$_2$ synthetic lines produced by the radiative transfer code. The best $^{34}$SO$_2$/$^{32}$SO$_2$ ratio derived is of 0.095$\pm$0.015 assuming an homogeneous distribution%(reduced $\chi^2$=1.13)%
, and 0.08$\pm$0.015 assuming the rescaled Feaga distribution%(reduced $\chi^2$=1.17)%. 
%Assuming the $^{33}$S/$^{34}$S ratio measured by \citet{howell1989}, and neglecting other S isotopes, this detection corresponds to a $^{34}$S isotopic abundance of 0.086$\pm$0.012 and 0.073$\pm$0.012 respectively.

\subsection{Tentative detection of KCl\label{KCl}}
\begin{figure}
\begin{center}
\begin{minipage}{10cm}
\includegraphics[width=9cm]{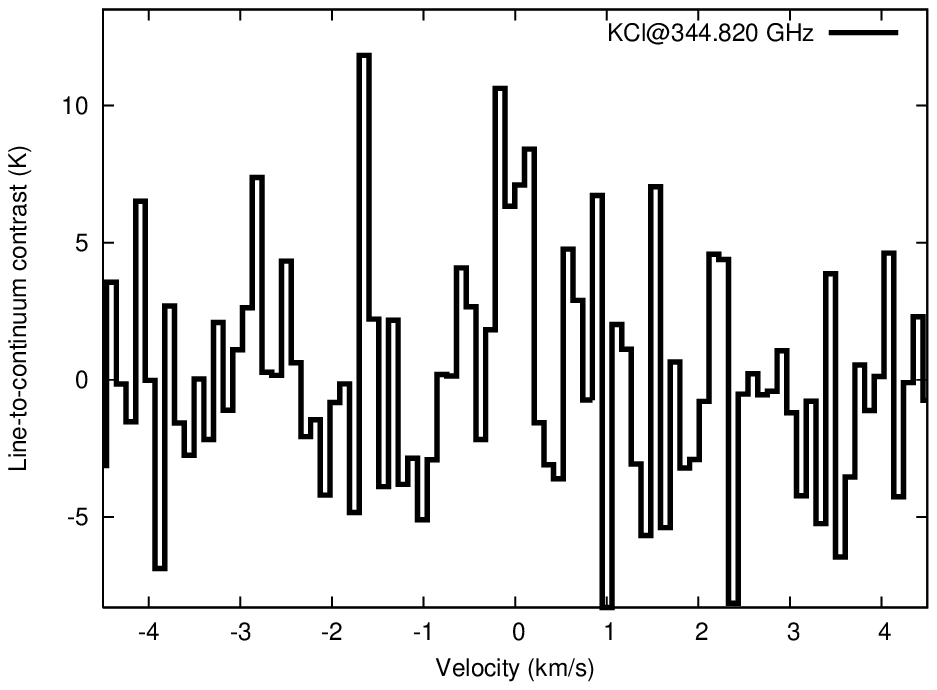}
\includegraphics[width=9cm]{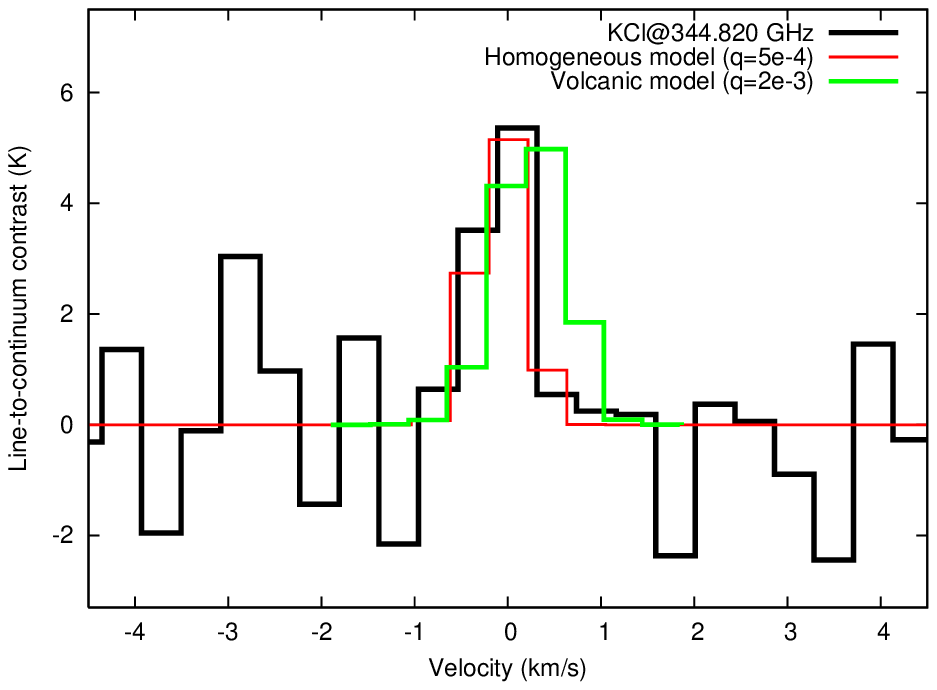}
\includegraphics[width=9cm]{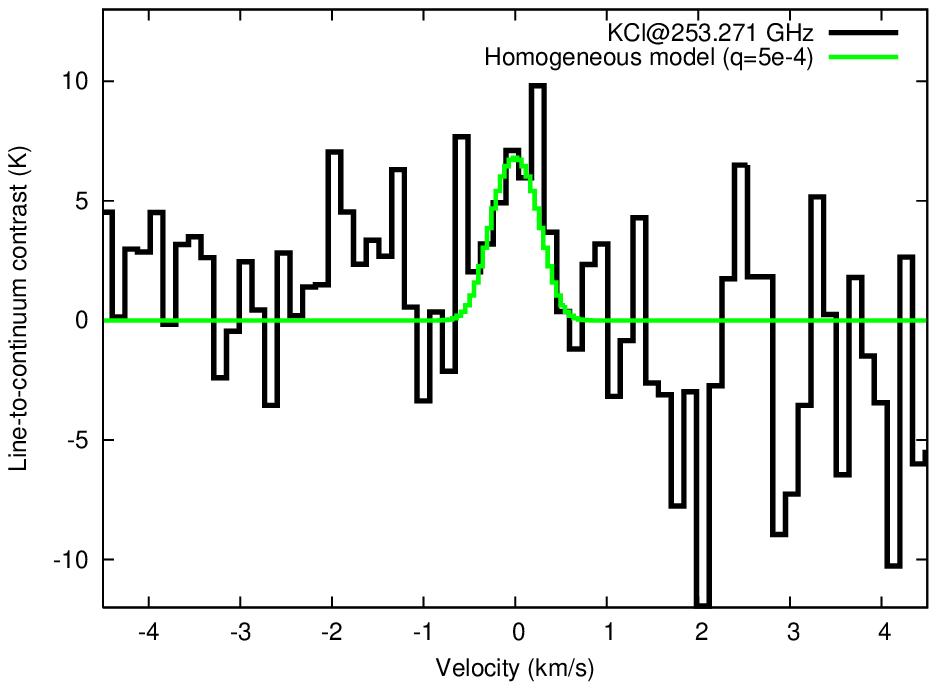}
\caption{Top: KCl line tentatively detected at APEX with the original spectral resolution. Middle: KCl line tentatively detected at APEX (with a spectral resolution degraded to 488~kHz), plotted against the best models for an homogeneous atmospheric distribution (q=5$\times$10$^{-4}$) and a purely volcanic model (q=2$\times$10$^{-3}$in plumes). Bottom: KCl line tentatively detected at IRAM-30m \citep{lellouch2003}, plotted against the homogeneous atmospheric distribution model\label{KClline}. }
\end{minipage}
\end{center}
\end{figure}

The 344.820~GHz KCl line is tentatively detected (Figure \ref{KClline}). At the original spectral resolution (top panel), a SNR per channel of just $\sim$2 is reached; however, when the data are spectrally binned down to a resolution of 488~kHz, corresponding to $\sim$half the linewidth (middle panel), a SNR of 3 per channel is reached. \\
Assuming that SO$_2$ and KCl are co-located (extended atmosphere model), the KCl line is well modeled using a KCl mixing ratio of (5$\pm$2)$\times$10$^{-4}$ assuming an homogeneous distribution and (4$\pm$2)$\times$10$^{-4}$ for a rescaled Feaga distribution, both corresponding to a KCl disk-averaged column density of $\sim$(3$\pm$1$)\times$10$^{12}$mol.cm$^{-2}$. \\
To explore the case of a purely volcanic origin for KCl, we derived the range of KCl mixing ratios in plumes that is compatible with the data, as explained in section \ref{volc}. In the most favorable situation where all known plumes are erupting (reference plume distribution), a KCl mixing ratio as low as 2$\times$10$^{-3}$ is sufficient. With such a distribution, the major part of the line emission is produced by the two large Pele-type Tvashtar and Pele plumes. Due to their geographic location (high northern latitudes on the leading hemisphere for Tvashtar, and as close as 15$^{\circ}$ from the western terminator on the trailing hemisphere for Pele), these plumes appear near the limbs, which is the geometry that most enhances the line emission. With a single erupting Pele plume near the sub-earth point (or 5 to 10 active Prometheus plumes distributed across each hemisphere), a KCl mixing ratio of 8$\times$10$^{-3}$ in plumes would be needed. With fewer active plumes, the observed contrast cannot be reached, as the KCl line saturates for mixing ratios higher than 8$\times$10$^{-3}$. \\ 
The extended atmospheric model proposed above also reproduces the contrast of the 253.271~GHz KCl line tentatively detected at the IRAM-30m antenna on January 2002 \citep[see Figure \ref{KClline}, bottom panel][]{lellouch2003}. Those observations were interpreted, based on the comparison with NaCl lines, as a maximum KCl/NaCl ratio of 1 for a NaCl mixing ratio of 2.5-5$\times$10$^{-4}$ (corresponding to a maximum disk-averaged KCl column density of 2.75-5.5$\times$10$^{12}$mol.cm$^{-2}$ assuming that all species are colocated). While the column density values appear to be consistent with the result obtained with our extended atmosphere model, a direct comparison of KCl mixing ratios cannot be drawn as the SO$_2$ atmospheric model used in \citet{lellouch2003} is quite different from our model; in particular it proposes a spatially concentrated and denser atmosphere to account for the large linewidths, as opposed to a global atmospheric distribution in our models. \\

\subsection{Upper limits on SiO and S$_2$O}
The search for the strong 347.330~GHz SiO line was unsuccessful. With a rms of 2~K on the spectrally binned data, the 3-$\sigma$ upper limit on the SiO mixing ratio is of $\sim$1.3$\times$10$^{-3}$ for an homogeneous distribution and $\sim$0.8$\times$10$^{-3}$ for a rescaled Feaga distribution, corresponding to an upper limit on the disk-averaged column density of $\sim$7$\times$10$^{12}$mol.cm$^{-2}$. \\
We used volcanic models to determine a range of SiO mixing ratios in plumes that would just reach the measured upper limit on the line contrast, assuming that SiO is purely maintained in the atmosphere by active plumes. In the most favorable case (reference plume distribution), the upper limit would be reached for an SiO mixing ratio in plumes of 6$\times$10$^{-3}$. With a single Pele plume near the sub-earth point, a mixing ratio of 2$\times$10$^{-2}$ would be required, and with a single Prometheus plume, a mixing ratio of 0.7.  \\ 
Finally, even when averaging the five spectral regions where S$_2$O transitions are expected within our observed spectral windows, S$_2$O could not be detected, only allowing us to a put a very loose mixing ratio upper limit of q$\sim$0.6.\\

\section{Discussion}

\subsection{Interpretation of the $^{34}$S/$^{32}$S isotopic ratio result}
The main observational result in this paper is the first reported detection of a $^{34}$S-bearing molecule in gas phase. The range of values that we derive for the $^{34}$S/$^{32}$S abundance ratio (within error bars, 0.065-0.120) is surprisingly higher than what was expected, as compared with the solar system reference $^{34}$S/$^{32}$S value of 0.044 \citep{lodders2003}. While \citet{howell1989} could not estimate the $^{34}$S/$^{32}$S abundance ratio, they detected a $^{33}$SO$_2$ band that allowed them to measure the $^{33}$S/$^{34}$S abundance ratio in the solid phase to be 0.13$\pm$0.07. This result is consistent with the solar system reference value \citep[0.18,][]{lodders2003}, suggesting that the global sulfur isotopic composition on Io may be similar to the solar system reference. \\
Our measurement of the $^{34}$S/$^{32}$S abundance ratio in the gas phase appears to be roughly twice as high as the solar system reference, as well as solar and ISM values \citep{asplund2009,lucas1998}. A $^{34}$S/$^{32}$S abundance ratio as high as our measurement has actually only been reported in redshifted absorption at $z\sim 0.89$ toward a quasar \citep{muller2006}.\\
\\
There are different ways to understand our result, depending on how and where fractionation occurs. First, the atmospheric $^{34}$S enrichment observed may directly reflect the sulfur isotopic composition in ionian magmas. This could be the same as the ionian bulk sulfur isotopic distribution, or result from fractionation occurring within the lava. $^{34}$S isotopic anomalies have been reported in terrestrial volcanic rocks, and are thought to be linked to variations in magma temperature and composition \citep{marini1994,sakai1982}. Anomalies have also been measured in chondrites, achondrite and iron meteorites \citep{gao1993,gao1993b,rai2005,gao1991}. However, the anomalies reported in terrestrial and meteoritical materials are at most a few per hundred. \\
Another possibility
% that could in addition possibly explain the presumed difference between our measurement and the results of \citet{howell1989}, 
is that mass-independent fractionation processes at work in the atmosphere significantly change the gas-phase isotopic composition with respect to the solid-phase isotopic distribution. SO$_2$ photolysis, chemical reactions and atmospheric transport have for example all been invoked to explain the enrichment in $^{34}$S observed in some terrestrial volcanic gases \citep{farquhar2001,baroni2007}, although the isotopic anomalies reported and modeled in those cases are of the order of a few per thousand only. \\
Finally, since the immediate source of SO$_2$ is dominated by sublimation equilibrium, it could be possible that sublimation/condensation cycles produce mass-dependent fractionation between $^{32}$SO$_2$ and $^{34}$SO$_2$, which are expected to have slight differences in equilibrium vapor pressure. These three possibilities may all be important to varying degree, but remain speculative and require quantitative evaluation to verify if they could account for the $^{34}$S enrichment that we measure. 
\\
\subsection{Potassium chloride}
The accumulation of tentative detections of KCl (sub)millimeter lines is an encouraging sign for definitive KCl detection using more sensitive instruments. The presence of K-bearing molecules in Io's atmosphere is strongly anticipated, as K atoms are suspected to be present in the upper atmosphere based on the spectral analysis of UV-airglows observed by Cassini \citep{geissler2004}, and have been detected in Io's corona \citep{brown2001}. In addition, K atoms and K-bearing dust have been detected respectively in Io's neutral clouds \citep{trafton1975} and in Jupiter's rings \citep{postberg2006}, that are both believed to be ultimately fed by Io's atmosphere. Using terrestrial volcanism analogies, \citet{fegley2000b}, \citet{moses2002}, \citet{schaefer2004} and \citet{schaefer2005} suggest that potassium is originally present in Io's erupting magma, from which it is easily volatilized due to the high temperatures of volcanic regions and low atmospheric pressures. KCl is the expected main K-carrier in these models and could be injected to the atmosphere directly from volcanic vents. Assuming Cl/S and K/S abundance ratios of 0.045 and 0.005, respectively, \citet{schaefer2005} predict a KCl mixing ratio in the plumes of $\sim$8$\times$10$^{-3}$ for a wide range of conduit pressures. They also predict that KCl should have a short lifetime in the atmosphere ($<$ 2 hours) due to condensation and photolysis. In regions where the bulk atmosphere is thinner, for example because of low sublimation-sustained SO$_2$ (nightside), gaseous K-bearing molecules could also be produced by sputtering of volcanic-ash coated surfaces by high-energy particles, in particular on the trailing hemisphere that is hit by plasma torus particles \citep{haff1981}. In this context, most of the KCl content on the dayside should correspond to a short-lived volcanically sustained atmosphere, indicative of on-going volcanic activity. \\
Our volcanic modeling results on KCl are consistent with this hypothesis. We have shown that assuming the KCl mixing ratio proposed by \citet{schaefer2005} ($\sim$8$\times$10$^{-3}$), a plausible distribution of active plumes could explain the tentatively detected KCl line. Our data are compatible with volcanism being the only KCl source on the dayside atmosphere. \\
\\
Considering the plume distribution models proposed in Section \ref{KCl} to explain the KCl tentative detection, we derive the corresponding total volcanically produced KCl. Given the SO$_2$ emission fluxes fixed by our individual plume models (1.1$\times$10$^{4}$~kg/s for a Prometheus plume, 2.2$\times$10$^{4}$~kg/s for a Pele plume), we estimate a production rate ranging between 7.8$\times$10$^{9}$ and 4$\times$10$^{10}$~KCl~molecules/s.cm$^{-2}$ (accounting for the emission from both hemispheres). We can compare this result to atmospheric escape rates measured at higher altitudes. \citet{mendillo2004} measured the Na escape rate during volcanically active periods to be between 1.8-5.6$\times$10$^{9}$~Na~atoms/s.cm$^{-2}$. Assuming the Na/K ratio measured in the corona by \citet {brown2001} (10$\pm$3), the K atoms escape rate inferred is of one to two orders of magnitude lower than our proposed volcanic KCl production rate. Our models are compatible with volcanically-produced KCl in Io's bulk atmosphere being the sole source of K-atoms for the corona. This result is consistent with the general behavior that atmospheric escape represents only a small fraction of the volcanic output, and in particular with the 50-100 times larger NaCl volcanic input compared to the upward flux of atomic Na and Cl at the top of the atmosphere \citep{moses2002,lellouch2003}. \\
%Assuming that the Na/K ratio in neutral clouds is the same as the ratio in the corona, the volcanic KCl production rate would also be higher than the replenishment rate required for a steady-state of the neutral K clouds \citep[$\sim$10$^{6}$,][]{brown1976}. Hence our models are compatible with volcanically-produced KCl in Io's bulk atmosphere being the sole source of K-atoms for the corona, and subsequently for neutral clouds.\\
%Under the same KCl abundance assumption, the data would also indicate that the two identified Pele-type plumes (Tvashtar and Pele) could not be both active at the time of the observations. Indeed, because of their location near the limbs, those two plumes would produce a line contrast about 50\% higher than what was detected.\\
\\
In addition to these results on the possible contribution of volcanism to atmospheric KCl content, our data allows us to constrain the Na/K atmospheric ratio. As explained in \citet{schaefer2004}, obtaining a stringent determination of the Na/K ratio in the bulk atmosphere can be a powerful tool to distinguish between different types of lavas (silicate, basalt, K-rich) and constrain vaporization temperature. In the lower atmosphere, where most of gaseous K and Na is believed to be carried in KCl and NaCl respectively \citep{fegley2000b, moses2002, schaefer2005}, this can be done by measuring the NaCl/KCl ratio. Our observed spectral windows did not include NaCl transitions that would allow a direct comparison. Using the 338.021~GHz NaCl line detected at the SMA \citep{moullet2010}, we derive disk-averaged NaCl column densities between 7-9$\times$10$^{-12}$mol.cm$^{-2}$, leading to a NaCl/KCl ratio of 2.7$^{+1.8}_{-1}$. \\
This result is consistent and close to the Na/K value proposed for the upper atmosphere by \citet{geissler2004} ($\sim$3.3). Both these measurements are much lower, by several orders of magnitude, than most Na/K ratios modeled for gases vaporized over lavas that could be Ionian analogues \citep{schaefer2004}, and significantly lower than cosmic and chondritic values \citep[16 and 15 respectively, ][]{asplund2009,lodders2003}. \citet{schaefer2004} address this issue in detail by proposing that, since Na is more volatile than K, fractional vaporization effects could over time lead to a depletion of the Na content in lavas and therefore in the atmosphere. Another proposition is the presence of originally K-rich lavas (ultra-potassic lavas with Na/K down to $\sim$2.8).
\\
We also note that our Na/K estimate is lower than the value measured in the corona based on Na and K atoms \citep[10$\pm$3,][]{brown2001}. This apparent Na/K ratio increase from the bulk atmosphere to Io's environment needs to be confirmed by higher-precision measurements, as it would be indicative of processes acting differently on K-bearing and Na-bearing molecules and atoms. \citet{brown2001} actually argue that their corona measurement should be considered as an upper limit on the Na/K ratio near the surface, as K atoms are more quickly photo-ionized than Na atoms.\\

\subsection{Silicon oxide}
The present understanding of Io's volcanism is that both silicate-based and sulfur-based magmas are driving volcanic activity \citep{williams2007}. The very high temperatures measured in some volcanic regions \citep{lopes1999} are much higher than the sulfur boiling point but consistent with silicate flows akin to those commonly found on Earth (basalts), or to ultramafic lavas \citep{mcewen1998}. The rigidity of the surface topographic features, including several km high mountains \citep{clow1980}, also points to a silicate-based crust that would explain Io's bulk density. While sulfur-bearing volatiles can be dissolved in silicate melts \citep{carr79}, the overall predominance of sulfur on Io's surface, atmosphere and environment, suggests the possibility of co-existent sulfur-based magmas. Most measured hot-spot temperatures are indeed low enough to be consistent with molten-sulfur magma \citep{mcewen1983}, and the morphology of some bright volcanic regions is indicative of sulfur flows \citep{mcewen2000}. No Si-carrying molecules have been positively identified on Io, and such a detection would provide an additional strong proof of the existence of silicate volcanism on Io.\\
The science case behind the interpretation of SiO content is similar to that of KCl. Vaporization models show that SiO should be the main Si-bearing volatile over silicate lavas at very high temperature \citep{schaefer2004}, and that it is expected to condense very quickly, so that its presence should be restricted to active volcanic plumes. However, unlike KCl, the amount of vaporized SiO depends strongly on lava temperature, with column densities above lavas expected to span four orders of magnitude (for a temperature range between 1700-2400~K). Based on the examples given in \citet{schaefer2004}, the SiO column density over a plume can vary between 2$\times$10$^{15}$mol.cm$^{-2}$ and 8$\times$10$^{18}$mol.cm$^{-2}$. Considering that the SO$_2$ column density over Pele and Prometheus plumes is of the order of 2$\times$10$^{18}$mol.cm$^{-2}$ based on plume-resolved measurements \citep{jessup2004,spencer2000}, this means a wide possible range for SiO mixing ratio in plumes from 1$\times$10$^{-3}$ to 4.\\
Our modeling shows that it is possible to reproduce our measured upper limit on the SiO line for a variety of plausible plume distributions, assuming mixing ratios in the range of 6$\times$10$^{-3}$ - 0.7, that are within the range of predicted SiO mixing ratios. Hence our upper limit measurement is entirely consistent with a purely volcanic source for SiO.\\
Using the same results presented above on NaCl content, and given that SiO is expected to be the main Si-carrier, we derive an upper limit on the Si/Na ratio of $\sim$1. This value is much less stringent than the value obtained by \citet{na1998} on the Si/Na ratio in the corona (0.014). Our upper limit also appears to be consistent with the Si/Na ratio measured in the vaporized phase above most terrestrial basalts \citep[of the order of 1$\times$10$^{-3}$-1$\times$10$^{-6}$][]{schaefer2004}, but is too imprecise to constrain lava composition.\\
\\
\section{Conclusions}
This work demonstrates how submillimeter observations can be used to derive unique constraints on the composition and sources of Io's lower atmosphere, which are essential to understand Io's volcanism and atmospheric processes, but also demonstrates the sensitivity limits reached by APEX observations that prevent us from performing spectroscopic searches deep enough to detect further minor species. %Observations using the EMIR receiver at the IRAM-30 m, while offering a much larger collecting area, are hampered by the lower atmospheric quality of the site, and hence could only improve upon detection limits by a factor $\sim$3 (for a similar integration time), preventing us from performing deep enough chemical searches.%
In particular, obtaining better upper limits on volcanic tracers such as SiO would strongly constrain the characterization of ionian volcanic regimes. In addition, the interpretation of disk-averaged spectra of a strongly spatially variable atmosphere is inherently uncertain, and in particular for optically thick lines, the necessary assumptions on spatial distribution can lead to significantly different results in column density. With an apparent size of 0.9-1.2", the use of interferometric facilities is necessary to resolve Io's disk, and observations with the SMA and IRAM-PdBI offer a spatial resolution down to 0.4" allowing for a first-order assessment of atmospheric coverage.\\
The Atacama Large Millimeter Array (ALMA), a 64-element array that is approaching completion near the APEX site, will provide the necessary boosts in sensitivity and spatial resolution. Due to its very large collecting area and state-of the art receivers, detection limits on minor species could be lowered by a factor of $\sim$50-70, allowing for searches of more species and isotopes. Extended array configurations could be used to reach spatial resolution down to 0.15" (corresponding to $\sim$400-500~km, i.e., the size of a Prometheus-type plume) with a sufficient SNR at $\sim$350~GHz, allowing direct determination of the local SO$_2$ column density as well as helping to establish the link between atmospheric species and their respective sources. ALMA observations are then one of the most promising prospects in the next decade for the advancement of the understanding of Io at large.

\acknowledgements{This paper is based on observations obtained at the Atacama Pathfinder EXperiment (APEX) telescope. APEX is a collaboration between the Max Planck Institute for Radio Astronomy, the European Southern Observatory, and the Onsala Space Observatory. We are grateful to the APEX staff for scheduling this challenging observing program, and in particular to M. Dumke for developing dedicated observation software. We thank L. Feaga and C. Moore who kindly shared their atmospheric and volcanic models, which were used for the analysis presented in the paper. A.M. is a Jansky Fellow at the National Radio Astronomy Observatory, a facility of the National Science Foundation operated under cooperative agreement by Associated Universities, Inc.}


\bibliographystyle{aa}
\bibliography{biblio}

%\end{thebibliography}

\end{document}